\documentclass[12pt]{article}
\usepackage{a4wide}
\usepackage{epsfig}
\usepackage{amsmath}

\newlength{\absize}
\catcode`@=11
\def\citer{\@ifnextchar [{\@tempswatrue\@citexr}{\@tempswafalse\@citexr[]}}

\def\@citexr[#1]#2{\if@filesw\immediate
  \write\@auxout{\string\citation{#2}}\fi
  \def\@citea{}\@cite{\@for\@citeb:=#2\do
    {\@citea\def\@citea{--\penalty\@m}\@ifundefined
       {b@\@citeb}{{\bf ?}\@warning
       {Citation `\@citeb' on page \thepage \space undefined}}%
\hbox{\csname b@\@citeb\endcsname}}}{#1}}
\catcode`@=12

\begin{document}
  \thispagestyle{empty}
  \pagestyle{empty}
  \renewcommand{\thefootnote}{\fnsymbol{footnote}}
\newpage\normalsize
    \pagestyle{plain}
    \setlength{\baselineskip}{4ex}\par
    \setcounter{footnote}{0}
    \renewcommand{\thefootnote}{\arabic{footnote}}
\newcommand{\preprint}[1]{%
  \begin{flushright}
    \setlength{\baselineskip}{3ex} #1
  \end{flushright}}
\renewcommand{\title}[1]{%
  \begin{center}
    \LARGE #1
  \end{center}\par}
\renewcommand{\author}[1]{%
  \vspace{2ex}
  {\Large
   \begin{center}
     \setlength{\baselineskip}{3ex} #1 \par
   \end{center}}}
\renewcommand{\thanks}[1]{\footnote{#1}}
\begin{flushright}
\end{flushright}
\vskip 0.5cm

\begin{center}
{\large \bf Deformed Two-Photon Squeezed States in Noncommutative
Space}
\end{center}
\vspace{1cm}
\begin{center}
Jian-zu Zhang$^{\;\ast}$
\end{center}
\vspace{1cm}
\begin{center}
Institute for Theoretical Physics, Box 316, East China University
of Science and Technology, Shanghai 200237, P. R. China
\end{center}
\vspace{1cm}

\begin{abstract}
Recent studies on non-perturbation aspects of noncommutative
quantum mechanics explored a new type of boson commutation
relations at the deformed level, described by deformed
annihilation-creation operators in noncommutative space. This
correlated boson commutator correlates different degrees of
freedom, and shows an essential influence on dynamics.
This paper devotes to the development of formalism of deformed
two-photon squeezed states in noncommutative space. General
representations of deformed annihilation-creation operators and
the consistency condition
for the electromagnetic wave with a single mode of frequency in NC
space are obtained. Two-photon squeezed states are studied. One
finds that variances of the dimensionless hermitian quadratures of
the annihilation operator in one degree of freedom include
variances in the other degree of freedom. Such correlations show
the new feature of spatial noncommutativity and allow a deeper
understanding of the correlated boson commutator.

\end{abstract}

\begin{flushleft}
${^\ast}$ E-mail address: jzzhang@ecust.edu.cn

\end{flushleft}
\clearpage
In the investigation of physics in noncommutative(NC) space
\citer{CDS,SW} motivated by arguments of string theory, apart from
studies of field theory, noncommutative quantum mechanics(NCQM)
has recently attracted some attention \citer{CST,JZZ04b}. As the
one-particle sector of NC quantum field theory, studies of NCQM
show special meaning. It may clarify some phenomenological
consequences in solvable models. NC effects will only become
apparent as the NC high energy scale is approached. But it is
expected that, because of the incomplete decoupling mechanism
between the high energy sector and the low energy sector,
 there should be some relics of NC effects appear in the low energy
sector.
Studies of NCQM may explore such low energy relics of effects of
spatial noncommutativity.
In literature perturbation aspects of NCQM have been studied in
detail. The perturbation
approach is based on the Weyl-Moyal correspondence \citer{AW,RS},
according to which the usual product of functions should be
replaced by the star-product.
Because of the exponential differential factor in the
Weyl-Moyal product the non-perturbation treatment is difficulty.

Recent studies on non-perturbation aspects of NCQM clarified
\citer{JZZ04a} that in order to maintain Bose-Einstein statistics
at the non-perturbation level described by deformed
annihilation-creation operators in NC space, when the state vector
space of identical bosons is constructed by generalizing
one-particle quantum mechanics, the consistent ansatz of
commutation relations of the phase space variables should
simultaneously include space-space noncommutativity and
momentum-momentum noncommutativity. A new type of boson
commutation relations at the deformed level, called the correlated
boson commutator, was explored. The correlated boson commutator
correlates different degrees of freedom, and shows an essential
influence on dynamics. For example, it explored that the spectrum
of the angular momentum possesses, because of such correlating
effects, fractional eigenvalues \cite{JZZ04a}.

In literature many interesting topics of NC quantum theories have
been extensively investigated, from the Aharonov-Bohm effect to
the quantum Hall effect \citer{CDPST00,FGLMR}. In order to further
explore the influence of the correlated boson commutator on
dynamics, in this paper we study deformed two-photon squeezed
states.

In commutative space the boson algebra shows that
annihilation-creation operators in one degree of freedom are
independent of ones in the other degree of freedom. In NC space
new future appears, the correlated boson commutator shows that
there is a correlation between different degrees of freedom. In
order to explore the correlated effects this paper devotes to the
development of formalism of deformed two-photon squeezed states in
noncommutative space. It turns out that the variances of the
dimensionless hermitian quadratures of the annihilation operator
in one degree of freedom include the ones in the other degree of
freedom. These results show the new future of spatial
noncommutativity and allow a deeper understanding of the
correlated boson commutator.

In the following we first review the background
\cite{NP,JZZ04a,DN}.

In order to develop the NCQM formulation we need to specify the
phase space and the Hilbert space on which operators act. The
Hilbert space can consistently be taken to be exactly the same as
the Hilbert space of the corresponding commutative system
\citer{CST}.

As for the phase space we consider both space-space
noncommutativity (space-time noncommutativity is not considered)
and momentum-momentum noncommutativity. There are different types
of NC theories, for example, see a review paper \cite{DN}.

In the case of simultaneously space-space
 noncommutativity and momentum-momentum noncommutativity the
consistent NCQM algebra are:
\begin{equation}
\label{Eq:1e} [\hat x_{i},\hat
x_{j}]=i\xi^2\theta\epsilon_{ij}, \qquad [\hat x_{i},\hat
p_{j}]=i\hbar\delta_{ij}, \qquad [\hat p_{i},\hat
p_{j}]=i\xi^2\eta\epsilon_{ij},\;(i,j=1,2)
\end{equation}
where $\theta$ and $\eta$ are the constant parameters, independent
of position and momentum; their dimensions are, respectively,
$L^2$ and $M^2 L^2 T^{-2},$ where  $M, L, T$ are, respectively,
 dimensions of mass, length and time.
$\epsilon_{ij}$ is an antisymmetric unit tensor,
$\epsilon_{12}=-\epsilon_{21}=1,$ $\epsilon_{11}=\epsilon_{22}=0.$
$\xi=(1+\theta\eta/4\hbar^2)^{-1/2}.$
When $\eta=0,$ we have $\xi=1,$ the NCQM
algebra (\ref{Eq:1e}) reduces to the one which is extensively
discussed in literature for the case that only space-space are
noncommuting.

There are different ways to construct the creation-annihilation
operators. In order to explore NC effects at the non-perturbation
level
for our purpose we consider a system with a single mode of
frequency $\omega$, for example, the plain electromagnetic wave of
single frequency. For the dimension $i$ we find the general
representations of the deformed annihilation-creation operators
$\hat a_i$, $\hat a_i^\dagger$ $(i=1,2)$
\begin{equation}
\label{Eq:2e} \hat a_i=\frac{\omega}{\sqrt{2}c}\left (\hat
x_i +i\frac{c^2}{\hbar\omega^2}\hat p_i\right), \quad
\hat a_i^\dagger=\frac{\omega}{\sqrt{2}c}\left (\hat
x_i -i\frac{c^2}{\hbar\omega^2}\hat p_i\right).
\end{equation}
($c$ is speed of light in vacuum.) We notice that
Eqs.~(\ref{Eq:2e}) are $\omega$ dependent. Their NC parameter
dependent structures are different from the ones of two
dimensional harmonic oscillator in Ref.~\cite{JZZ04a}.

When the state vector space of identical bosons is constructed by
generalizing one-particle quantum mechanics, in order to maintain
Bose-Einstein statistics at the level described by $\hat
a_i^\dagger$, the basic assumption is that operators $\hat
a_i^\dagger$ and $\hat a_j^\dagger$ should be commuting. This
requirement leads to a consistency condition of NCQM algebra
\cite{JZZ04a}
\begin{equation}
\label{Eq:3e} \eta=\hbar^2\omega^4 c^{-4} \theta.
\end{equation}

{\bf Deformed Boson Algebra.} From Eqs.~(\ref{Eq:1e}) -
(\ref{Eq:3e}) it follows that the deformed boson algebra of $\hat
a_i$ and $\hat a_j^\dagger$ read \cite{JZZ04a}
\begin{equation}
\label{Eq:4e} [\hat a_i,\hat a_j^\dagger]=\delta_{ij}+
i\xi^2\omega^2 c^{-2} \theta,
\quad
[\hat a_i,\hat a_j]=0,\;(i,j=1,2)
\end{equation}
For the case $i=j$, Eqs.~(\ref{Eq:4e}) is the same commutation
relations as the ones in commutative space. This confirms that for
the same degree of freedom $i$ the operators
 $\hat a_i,\hat a_i^\dagger$ are the correct deformed
annihilation-creation operators.
For the case $i\ne j$, a new type of deformed commutation
relations between $\hat a_i$ and $\hat a_j^\dagger,$
called the correlated boson commutator, emerges,
\begin{equation}
\label{Eq:5e}
[\hat a_1,\hat
a_2^\dagger] =i\xi^2\omega^2 c^{-2} \theta,
\end{equation}
which correlates different degrees of freedom.
At the level of deformed operators the effect of spatial
noncommutativity are coded in Eq.~(\ref{Eq:5e}).

It is worth noting that Eq.~(\ref{Eq:5e}) is
consistent with {\it all} principles of
quantum mechanics and Bose-Einstein statistics.

The NCQM algebra (\ref{Eq:1e})
has different possible perturbation realizations \cite{NP}. To the
linear terms of phase space variables in commutative space,
the ansatz of the perturbation expansions of $\hat x_{i}$ and
$\hat p_{i}$, consistent with NCQM algebra (\ref{Eq:1e}), is
\begin{equation}
\label{Eq:6e} \hat
x_{i}=\xi[x_{i}-\frac{1}{2\hbar}\theta\epsilon_{ij}p_{j}], \quad
\hat p_{i}=\xi[p_{i}+\frac{1}{2}\hbar\omega^4 c^{-4}\theta
\epsilon_{ij}x_{j}].
\end{equation}
where $(x_i, p_i)$ are the phase space variables in commutative space,
and $[x_{i},x_{j}]=[p_{i},p_{j}]=0,
[x_{i},p_{j}]=i\hbar\delta_{ij}.$

The ansatz of the perturbation expansions of $\hat a_i$ and $\hat
a_i^\dagger$ is:
\begin{equation}
\label{Eq:7e} \hat a_{i}=\xi[a_{i}+\frac{i}{2}
\omega^2 c^{-2}\theta
\epsilon_{ij}a_j],\quad
\hat a_{i}^\dagger=\xi[a_{i}-\frac{i}{2}\omega^2 c^{-2}\theta
\epsilon_{ij}a_j],
\end{equation}
where  $(a_i, a_i^\dagger)$ are annihilation-creation operators
in commutative space.
The relations between $(a_i, a_i^\dagger)$ and $(x_i, p_i)$ are
$x_i=\sqrt{\hbar/(2\mu\omega})\left (a_i+a_i^\dagger\right)$,
$p_i=-i\sqrt{\mu\omega\hbar/2}\left (a_i -a_i^\dagger\right).$
We have $[a_{i},a_{j}]=[a_i^\dagger,a_j^\dagger]=0,
[a_{i},a^{\dagger}_{j}]=i\delta_{ij}.$ Eq.~(\ref{Eq:7e}) is
consistent with the deformed boson algebra (\ref{Eq:4e}),
specially including correlating effects of Eq.~(\ref{Eq:5e}).

In summary, structures of the operators $(\hat a_i, \hat
a_i^\dagger)$, the consistency condition and the perturbation
expansions of $(\hat x_i, \hat p_i)$ and $(\hat a_i$, $\hat
a_i^\dagger)$ are determined by the characteristic parameters of
the system. The NC parameter dependent structures of
Eqs.~(\ref{Eq:2e}), (\ref{Eq:3e}), (\ref{Eq:5e}) - (\ref{Eq:7e})
are different from the ones of two dimensional harmonic oscillator
in Ref.~\cite{JZZ04a}.

Now we investigate the influence of the correlated boson
commutator Eq.~(\ref{Eq:5e}) on deformed two-photon squeezed
states in NC space.

In discussions of the Heisenberg minimal uncertainty relation
special attention has been focused on coherent states and
squeezed states of the light field \cite{Walls}.
The idea of squeezing has both fundamental and practical
interests.
Here we consider a special squeezed state suggested in
Ref.~\cite{Janszky}.
Such kind of squeezed states is easy to generalize to the case in
NC space.

In commutative space a coherent state is defined as an eigenstate
of the annihilation operator $a$,
$a|\alpha\rangle=\alpha|\alpha\rangle$ with a complex eigenvalue
$\alpha$. The coherent state $|\alpha\rangle$ is represented as
\begin{equation} 
\label{Eq:8e} |\alpha\rangle=N_\alpha\left[\sum_{n=0}^\infty
\frac{\alpha^n} {\sqrt{n!}}|n\rangle\right] =N_\alpha exp(\alpha
a^\dagger)|0\rangle,
\end{equation}
where $|n\rangle$ is the number states, $N|n\rangle=n|n\rangle,$ $
N=a^\dagger a$ is the number operator. From Eq.~(\ref{Eq:8e})
$\langle\alpha|-\alpha\rangle=N_{-\alpha}^*N_\alpha\,
exp(-\alpha^2)$. For simplicity we may choose the phase factor so
that $\alpha$ is real, thus in (\ref{Eq:8e}) the normalization
constant is $N_\alpha=exp(-\alpha^2/2)$.

The special squeezed state considered in Ref. \cite{Janszky} is as
follows.
An effective squeezing can be achieved by superposition of coherent
states along a straight line on the $\alpha$ plane.
This mechanism opens new possibility for squeezing, e.g., of the
molecular vibrations during a Franck-Condon transition induced by a short
coherent light pulse.
For a single mode of frequency $\omega$ the electric field operator
$E(t)$ is represented as
$E(t)=E_0[a\exp(-i\omega t)+a^\dagger\exp(i\omega t)]$,
where $a$ and $a^\dagger$ are the
annihilation and creation operators of photon field.
This squeezed state is defined as
\begin{equation} 
\label{Eq:9e} |\alpha,\pm\rangle
=c_\pm\left(|\alpha\rangle\pm|-\alpha\rangle\right),
\end{equation}
which satisfy
$a|\alpha,\pm\rangle=\alpha c_\pm c_\mp^{-1}|\alpha,\mp\rangle$,
$a^2|\alpha,\pm\rangle=\alpha^2|\alpha,\pm\rangle$, and
$\langle\alpha,\pm|\alpha,\mp\rangle=0$.
The normalization constants are
$c_\pm^2=exp(\alpha^2)/\{2[exp(\alpha^2)\pm exp(-\alpha^2)]\}$.

In the state $|\alpha,+\rangle$ squeezing appears.
Let $X$ and $Y$ be the dimensionless hermitian quadratures of the
annihilation operator: $a=X+iY$.
The variances of $X$ and
$Y$ in this state are \cite{Janszky}:
\begin{equation} 
\label{Eq:10e} (\Delta X)^2 =\frac{1}{4}+\frac{\alpha^2 exp(
\alpha^2)} {exp(\alpha^2)+exp(-\alpha^2)},\quad
(\Delta Y)^2 = \frac{1}{4}-\frac{\alpha^2 exp(-\alpha^2)}
{exp(\alpha^2)+exp(-\alpha^2)}.
\end{equation}
Here for any normalized state $|\psi\rangle,$
the variances of an operator $F$ is defined as
$\Delta F\equiv \left[\left(\psi,\left(F-
\bar F\right)^2
\psi\right)\right]^{1/2},$  $\bar F\equiv (\psi,F\psi).$
It is noticed that the state $|\alpha,+\rangle$ is squeezed, i.e.,
the variance $(\Delta Y)^2$ is less than Heisenberg's
minimal uncertainty 0.25.
The maximum squeezing appears at $\alpha_0^2=0.64$,
where $(\Delta Y)_0^2=0.111$.
Beyond $\alpha_0^2$, as $\alpha^2$ increases,
$(\Delta Y)^2$ monotonically increases to Heisenberg's
minimal uncertainty 0.25.

{\bf Two-Photon Squeezed States.} The correlated boson commutator
(\ref{Eq:5e}) shows that in NC space there is a correlation
between different degrees of freedom. Now we clarify the influence
(\ref{Eq:5e}) on two-photon squeezed states. For this purpose we
first clarify the situation of two-photon squeezed states in
commutative space.

From two-photon coherent state $|\alpha_1\rangle|\alpha_2\rangle$,
$a_i|\alpha_1\rangle|\alpha_2\rangle=
\alpha_i|\alpha_1\rangle|\alpha_2\rangle$, (i=1, 2)
we can construct three types of two-photon squeezed states:
\begin{eqnarray}
\label{Eq:i-ii-iii} &&(i)\; |I,\pm\rangle\equiv
|\alpha_1,\pm\rangle|\alpha_2\rangle,
\nonumber\\
&&(ii)\; |II,\pm\rangle\equiv |\alpha_1,\pm\rangle
|\alpha_2,\pm\rangle,
\nonumber\\
&&(iii)\; |III,\pm\rangle
=c_{3\pm}\left(|\alpha_1,\alpha_2\rangle\pm|
-\alpha_1,-\alpha_2\rangle\right),
\end{eqnarray}
From the definitions the states $|I,\pm\rangle$ and
$|II,\pm\rangle$ are normalized. For the states $|III,\pm\rangle$
the normalization constants
$c_{3\pm}^2=exp(\alpha_1^2+\alpha_2^2)/
\{2[exp(\alpha_1^2+\alpha_2^2)\pm exp(-\alpha_1^2-\alpha_2^2)]\}$.

For the degree of freedom $i$ let $a_i=X_i+iY_i$. Where the
variables $X_{i}$ and  $Y_{i}$ satisfy $[X_{i},
Y_{j}]=i\frac{1}{2}\delta_{ij}.$ In the state $|I,+\rangle$ the
variances of $X_1$ and $Y_1$ of the photon 1  are the same as for
a single squeezed state represented by Eq.~(\ref{Eq:10e}):
$(\Delta X_1)_{I+}^2 = \frac{1}{4}+\alpha_1^2 exp(\alpha_1^2)
/[exp(\alpha_1^2)+exp(-\alpha_1^2)]$,
$(\Delta Y_1)_{I+}^2 = \frac{1}{4}-\alpha_1^2 exp(-\alpha_1^2)
/[exp(\alpha_1^2)+exp(-\alpha_1^2)]$;
The photon 2 is
a 'spectator' in a coherent state, the variances of $X_2$ and
$Y_2$ are:
$(\Delta X_2)_{I+}^2 = (\Delta Y_2)_{I+}^2 = \frac{1}{4}$.
 In the states $|II,+\rangle$ the photons 1 and 2
are independently squeezed:
$(\Delta X_1)_{II+}^2 = (\Delta X_1)_{I+}^2 $,
$(\Delta Y_1)_{II+}^2 = (\Delta Y_1)_{I+}^2 $;
Changing the sub-index $1\to 2$, we obtain $(\Delta X_2)_{II+}^2$
and $(\Delta Y_2)_{II+}^2$.
In the state $|III,+\rangle$ the variances
of $X_1$ and $Y_1$ of the photon 1  are
\begin{subequations}
\begin{eqnarray} 
\label{Eq:12ae}
(\Delta X_1)_{III+}^2 &=& \frac{1}{4}+
\frac{\alpha_1^2 exp(\alpha_1^2+\alpha_2^2)}
{exp(\alpha_1^2+\alpha_2^2)+exp[-(\alpha_1^2+\alpha_2^2)]},
\end{eqnarray}
\begin{eqnarray}
\label{Eq:12be}
(\Delta Y_1)_{III+}^2 &=& \frac{1}{4}-
\frac{\alpha_1^2 exp[-(\alpha_1^2+\alpha_2^2)]}
{exp(\alpha_1^2+\alpha_2^2)+exp[-(\alpha_1^2+\alpha_2^2)]}.
\end{eqnarray}
\end{subequations}
Changing the sub-index $1\to 2$, we obtain $(\Delta X_2)_{III+}^2$
and $(\Delta Y_2)_{III+}^2$.
In the following we only consider the states $|I,+\rangle$ and
$|III,+\rangle$.

{\bf Deformed Two-Photon Squeezed States in NC Space.} From
Eqs.~(\ref{Eq:6e}) and (\ref{Eq:7e}) the dimensionless hermitian
quadratures of the deformed annihilation operator $\hat a_i=\hat
X_i+i\hat Y_i$ are
\begin{equation}
\label{Eq:13e} \hat X_{i}=\xi[X_{i}-\frac{1}{2}\omega^2 c^{-2}
\theta\epsilon_{ij}Y_j],\quad
\hat Y_{i}=\xi\left[Y_{i}+\frac{1}{2}\omega^2 c^{-2}\theta
\epsilon_{ij}X_j\right].
\end{equation}
The variances of $\hat X_{i}$ and
$\hat Y_{i}$ in the state $|I,+\rangle$ are
\begin{equation} 
\label{Eq:14e}
(\Delta \hat X_1)_{I+}^2 =\xi^2\left[(\Delta X_1)_{I+}^2
+\frac{1}{16}\omega^4 c^{-4}\theta^2\right], \quad
(\Delta \hat Y_1)_{I+}^2 =\xi^2\left[(\Delta Y_1)_{I+}^2
+\frac{1}{16}\omega^4 c^{-4}\theta^2\right],
\end{equation}
\begin{equation}
\label{Eq:15e}
(\Delta \hat X_2)_{I+}^2 =\frac{1}{4}\xi^2\left[1
+\omega^4 c^{-4}\theta^2(\Delta Y_1)_{I+}^2\right],
\quad
(\Delta \hat Y_2)_{I+}^2 =\frac{1}{4}\xi^2\left[1
+\omega^4 c^{-4}\theta^2(\Delta X_1)_{I+}^2\right].
\end{equation}
The variances of $\hat X_{1}$ and
$\hat Y_{1}$ in the state $|III,+\rangle$ are
\begin{eqnarray} 
\label{Eq:16e}
(\Delta \hat X_1)_{III+}^2 =\xi^2\left[(\Delta X_1)_{III+}^2
+\frac{1}{4}\omega^4 c^{-4}\theta^2(\Delta Y_2)_{III+}^2\right],
\nonumber \\
(\Delta \hat Y_1)_{III+}^2 =\xi^2\left[(\Delta Y_1)_{III+}^2
+\frac{1}{4}\omega^4 c^{-4}\theta^2(\Delta X_2)_{III+}^2\right].
\end{eqnarray}
Changing the sub-index $1\to 2$, we obtain
$(\Delta \hat X_2)_{III+}^2$ and $(\Delta \hat Y_2)_{III+}^2$.

We notice that Eqs.~(\ref{Eq:15e}) and (\ref{Eq:16e}) show
correlating effects.
For example, the variance $(\Delta \hat X_1)_{III+}^2$ in the
degree of freedom 1 includes a $\theta$ dependent variance
$(\Delta Y_2)_{III+}^2$ in the degree of freedom 2, etc.

Now we estimate the contribution of the correlating effects. There
are different bounds of the scale of the parameter $\theta$ set by
experiments: the estimation from the Lorentz symmetry violation
\cite{CHKLO},  measurements of the Lamb shift \cite{CST} and
clock-comparison experiments \cite{MPR}. If we adopt the  bound of
the parameter $\theta/(\hbar c)^2\le (10^4 \;GeV)^{-2}
\cite{CST},$ and want the coefficient of the $\theta$ dependent
term $\frac{1}{4}\omega^4 c^{-4}\theta^2
=\frac{1}{4}(\hbar\omega)^4 (\theta/\hbar^2c^2)^2$ in
Eqs.~(\ref{Eq:15e}) and (\ref{Eq:16e}) to reach a value, say $\sim
10^{-4}$,
the corresponding frequency is $\nu \sim 10^{26} \;Hz$.

The correlated boson commutator, exposed in the maintenance of
Bose-Einstein statistics at the non-perturbation level described
by deformed annihilation-creation operators, is a clear indication
of NC effects.
Any physical observable can be formulated by a bilinear
representation of $\hat a_i$ and $\hat a_i^\dagger,$
thus the $\omega^4$ factor in the $\theta$ dependent term in
Eqs.~(\ref{Eq:15e}) and (\ref{Eq:16e}) is a general feature for
processes with a single mode of frequency in NC space. We conclude
that for exploring NC effects a process with a high frequency is
favorable.

\vspace{0.3cm}

This work has been supported by the National Natural Science
Foundation of China under the grant number 10074014 and by the
Shanghai Education Development Foundation.

\clearpage



\begin{thebibliography}{99}
\bibitem{CDS}
A. Connes, M. R. Douglas, A. Schwarz,
JHEP {\bf 9802}, 003 (1998) {\bf hep-th/9711162}.

\bibitem{DH}
M. R. Douglas, C. M. Hull,
JHEP {\bf 9802}, 008 (1998) {\bf hep-th/9711165}.

\bibitem{AAS}
F. Ardalan, H. Arfaei, M. M. Sheikh-Jabbari,
JHEP {\bf 9902}, 016 (1999) {\bf hep-th/9810072}.

\bibitem{CH99}
S-C. Chu, P-M. Ho,
Nucl. Phys. {\bf B550}, 151 (1999) {\bf hep-th/9812219}.

\bibitem{CH00}
S-C. Chu, P-M. Ho,
Nucl. Phys. {\bf B568}, 447 (2000) {\bf hep-th/9906192}.

\bibitem{Sch}
V. Schomerus,
JHEP {\bf 9906}, 030 (1999) {\bf hep-th/9903205}.

\bibitem{SW}
N. Seiberg and E. Witten, JHEP {\bf 9909}:032 (1999)
{\bf hep-th/9908142}.

\bibitem{CST}
M. Chaichian, M. M. Sheikh-Jabbari, A. Tureanu,
Phys. Rev. Lett. {\bf 86}, 2716 (2001) {\bf hep-th/0010175}.

\bibitem{GLR}
J. Gamboa, M. Loewe, J. C. Rojas,
Phys. Rev. {\bf D64}, 067901 (2001) {\bf hep-th/0010220}.

\bibitem{NP}
V. P. Nair, A. P. Polychronakos,
Phys. Lett. {\bf B505}, 267 (2001) {\bf hep-th/0011172}.

\bibitem{KD}
D. Kochan, M. Demetrian,
{\bf hep-th/0102050}

\bibitem{MP}
B. Morariu, A. P. Polychronakos,
Nucl. Phys. {\bf B610}, 531 (2001) {\bf hep-th/0102157}.

\bibitem{HS}
A. Hatzinikitas, I. Smyrnakis, J. Math. Phys. {\bf 43} 113 (2002)
{\bf hep-th/0103074}.

\bibitem{GLMR}
J. Gamboa, M. Loewe, F. Mendez, J. C. Rojas,
Int. J. Mod. Phys. {\bf A17}, 2555 (1999) {\bf hep-th/0106125}.

\bibitem{BNS}
S. Bellucci, A. Neressian, C. Sochichiu,
Phys. Lett. {\bf B522}, 345 (2001) {\bf hep-th/0106138}.

\bibitem{SS}
A. Smailagic, E. Spallucci,
Phys. Rev. {\bf D65}, 107701 (2002) {\bf hep-th/0108216}.

\bibitem{CS}
H. R. Christiansen, F. A. Schaposnik,
Phys. Rev. {\bf D65}, 086005 (2002) {\bf hep-th/0106181}.

\bibitem{A}
C. Acatrinei,
JHEP {\bf 0109}, 007 (2001) {\bf hep-th/0107078}.

\bibitem{HK}
P-M. Ho, H-C. Kao,
Phys. Rev. Lett. {\bf 88}, 151602 (2002) {\bf hep-th/0110191}.

\bibitem{JZZ04a}
Jian-zu Zhang,
Phys. Lett. {\bf B584}, 204 (2004) {\bf hep-th/0405135}.

\bibitem{JZZ04b}
Jian-zu Zhang, Testing Spatial Noncommutativity via Rydberg Atoms,
Phys. Rev. Lett. (in press) {\bf hep-ph/0405143}.

\bibitem{AW}
L. Alvarez-Gaume, S. R. Wadia,
Phys. Lett. {\bf B501}, 319 (2001) {\bf hep-th/0006219}.

\bibitem{MS}
A. Micu, M. M. Sheikh-Jabbari,
JHEP {\bf 0101}: 025 (2001) {\bf hep-th/0008057}.

\bibitem{RS}
I. F. Riad, M. M. Sheikh-Jabbari,
JHEP {\bf 0008}: 045 (2000) {\bf hep-th/0008132}.

\bibitem{CDPST00}
M. Chaichian, A. Demichev, P. Presnajder, M.M. Sheikh-Jabbari,
A. Tureanu
Phys. Lett. {\bf B527}, 149 (2002) {\bf hep-th/0012175}.

\bibitem{CDPST01}
M. Chaichian, A. Demichev, P. Presnajder, M.M. Sheikh-Jabbari,
A. Tureanu
Nucl. Phys. {\bf B611}, 383 (2001) {\bf hep-th/0101209}.

\bibitem{GR}
S. S. Gubser, M. Rangamani,
JHEP {\bf 0105}: 041 (2001) {\bf hep-th/0012155}.

\bibitem{Suss}
L. Susskind ,
{\bf hep-th/0101029}.

\bibitem{Poly}
A. P. Polychronakos ,
JHEP {\bf 0104}: 011 (2001) {\bf hep-th//0103013}.

\bibitem{HR}
S. Hellerman, M. V. Raamsdonk,
JHEP {\bf 0110}: 039 (2001) {\bf hep-th/0103179}.

\bibitem{FGLMR}
H. Falomir, J. Gamboa, M. Loewe, F. Mendez, J.C. Rojas,
Phy. Rev. {\bf D66}, 045018 (2002) {\bf hep-th/0203260}.

\bibitem{DN}
M. R. Douglas, N. A. Nekrasov,
Rev. Mod. Phys. {\bf 73}, 977 (2001) {\bf hep-th/0106048}.

\bibitem{Walls} 
D. F. Walls, Nature, 306 (1983) 141;
R. E. Slusher, L. W. Hollberg, B. Yurke, J. C. Mertz and J. F. Valley,
Phys. Rev. Lett.\ 55 (1985) 2409;
L.-A. Wu, H. J. Kimble, J. L. Hall, and H. Wu,
Phys. Rev. Lett. 57 (1986) 2520;
H. J. Kimble and D. F. Walls, J. Opt. Soc. Am. B4 (1987) 1450;
E. Giacobino and C. Fabre, Appl. Phys. B55 (1992) 189.

\bibitem{Janszky} 
J. Janszky and A. V. Vinogradov, Phys. Rev. Letters 64 (1990)
2771; and references therein.

\bibitem{CHKLO}
S. M. Carroll, J. A. Harvey, V. A. Kostelecky, C. D. Lan, T.
Okamoto, Phys. Rev. Lett. {\bf 87}, 141601 (2001) {\bf
hep-th/0105082}.

\bibitem{MPR}
I. Mocioiu, M. Pospelov, R. Roiban, Phys. Lett. {\bf B489}, 390 (2000)
{\bf hep-th/00005191}.



\end{thebibliography}
\end{document}